\documentstyle[preprint,aps]{revtex}

\newcommand {\be}{\begin{equation}}
\newcommand {\ee}{\end{equation}}

\begin{document}
\draft

\title{Disturbance Propagation in Chaotic Extended Systems
\\with Long-Range Coupling}
\vskip 0.3 cm
\author{Alessandro  Torcini$^{(1)}$ and Stefano Lepri$^{(2,3)}$ \\
\vskip 0.4 cm
(1) {\it Theoretische Physik, Bergische Universit\"at-Gesamthochschule
Wuppertal,\\ D-42097 Wuppertal, Germany}\\
\vskip 0.3 cm
(2) {\it Dipartimento di Fisica dell' Universit\`a 
and Istituto Nazionale di Fisica Nucleare,\\
I-40127 Bologna, Italy}\\
\vskip 0.3 cm
(3) {\it Istituto Nazionale di Ottica,
I-50125 Firenze, Italy}\\
}
\vskip 0.3 cm
\date{\today}
 
\maketitle
\begin{abstract}
Propagation of initially localized perturbations is investigated in 
chaotic coupled map lattices with long-range couplings decaying as a 
power of the distance. The initial perturbation propagates
exponentially fast along the lattice, with a rate given by the ratio 
of the maximal Lyapunov exponent and the power of the coupling.
A complementary description in terms of a suitable comoving 
Lyapunov exponent is also given. 
\end{abstract}

\vskip 0.9 cm
\pacs{PACS numbers: \ 05.45.+b}

Propagation of fronts in spatially extended systems is
a topic of wide interest, in several scientific contexts 
including fluid dynamics, dendritic growth, directional solidification,
liquid crystals, chemical reactions, flame propagation and epidemics
spreading \cite{cross,eckmann,mollison,kessler}.
In one spatial dimension, partial differential equations \cite{eckmann}, 
coupled map lattices (CML) \cite{cml} and cellular automata \cite{wolfram}
models have been studied. The fronts were either separating 
stable and unstable steady states \cite{piskunov,vansaar,others}, 
or regular and chaotic regions \cite{kan1,pikov}. 

A similar phenomenon is the propagation of disturbances in fully chaotic 
states of CMLs \cite{nonlin,disturb} and of the complex Ginzburg-Landau
equation \cite{cgle}. When the spatial coupling is local, 
e.g. for CML with nearest-neighbour interactions, two distinct regimes 
have been found to characterize the dynamics of fronts \cite{nonlin}. 
In the first 
regime, the velocity of the propagating front is calculated in the framework of 
a linear analysis, while in the second the value of the velocity 
is determinated by the full nonlinear evolution of the system. For a particular 
class of models, a transition between the two regimes has been observed when a 
parameter is varied \cite{nonlin}.  
The scenario is not expected to be different for couplings extending to more
than two neighbours, since this affects only the limit velocity of 
propagation \cite{chrono}.

If more general spatial couplings are considered, even the very definition of a 
velocity can be nontrivial. For example, epidemic models in one dimension 
exhibit a finite propagation speed if the spatial coupling (i.e. the infection
rate) decays exponentially or faster \cite{mollison}. The same result for some CML 
models is reported in Ref. \cite{vulp}, where it is observed that the time 
for a localized disturbance to overcome a threshold value at a certain distance $l$ 
from the initially perturbed site grows linearly with $l$ if the coupling is local 
or exponentially decaying, while is almost independent of $l$ for couplings decaying 
with power laws (with exponent not too large), indicating that the velocity is 
``infinite''. 

This Letter focuses on the study of disturbance propagation 
in systems with long range coupling, whose strength 
decays as a power law in space. 
The spatio-temporal evolution of an initially localized 
perturbation of a chaotic state is studied theoretically
and numerically. The perturbation is found to spread
exponentially fast along the lattice and an expression
for the corresponding rate is given. A comoving Lyapunov 
analysis confirms that the prediction is indeed correct.

In order to mimic a spatially extended system with long range interaction,
we introduce the CML model \cite{vulp} 
\be
u(x,t+1) = f( {\tilde u}(x,t))
\ee
where the indices $x$ labels the sites of a chain of length $L$ and
$t$ is the discrete time variable. The function $f(x)$ is a chaotic map of 
the interval and
\be
{\tilde u}(x,t) =  (1-\varepsilon) u(x,t) +  \varepsilon 
\sum_{\ 0<|x-y|\le\Delta_L} w(x-y) u(y,t) \quad ,
\label{cml_pow}
\ee
where $\Delta_L={L-1\over 2}$ ($L$ is assumed to be odd). As usual, periodic 
boundary conditions are considered $u(x,t)=u(x\pm L,t)$. The coupling constant 
$\varepsilon$ is bounded between 0 and 1, and we consider the coupling strength 
to decay in space as 
\be
w(x) = { c(\alpha) \over |x|^\alpha} 
\label{power_law}
\ee
where $\alpha>1$ to insure that the normalization constant
$c(\alpha)= [\sum_y |y|^{-\alpha}]^{-1}$ is bounded. 
Obviously, this constant is independent of $L$ in the thermodynamic
limit $L \to \infty$. For $\alpha\to +\infty$ the model (\ref{cml_pow}) 
reduces to the usual nearest-neighbours CML \cite{cml}. Notice that, at 
variance with the model of Ref.~\cite{vulp}, the long range coupling
is not introduced as a perturbation of the nearest-neighbours CML.

To study the propagation of localized disturbances in 
the system (\ref{cml_pow}), 
let us consider two chaotic trajectores $\{u(x,t)\}$ and $\{ v(x,t) \}$ 
generated by starting initial conditions which differ only around a single site 
$x=0$. More precisely, we assume that $u(x,0)$ is a typical chaotic state, 
obtained after all transients have died out, and $v(x,0)=u(x,0)+u_0(x)$,
where $u_0(x)$ is a function localized around the origin.

If only {\it linear} mechanisms of propagation are present 
\cite{nonlin,disturb}, we can assume that the evolution of 
$\delta u(x,t)=u(x,t) - v(x,t)$ is ruled by the linearized 
dynamics 
\be
\delta u(x,t+1) =  m(x,t)\left[
(1-\varepsilon) \delta u(x,t) + \varepsilon 
\sum_{0<|x-y|\le\Delta_L} w(x-y) \delta u(y,t) \right]\quad ,
\label{linmap}
\ee
where $m(x,t)=f^{\prime}( {\tilde u}(x,t))$ is the local multiplier 
along the assigned trajectory. This hypotesis is justified at least for 
large distances ($|x|\gg 1$) where the disturbance is small.
Moreover, its validity has also been numerically checked by comparing 
the evolution given by Eq.~(\ref{linmap}) with that of $u(x,t) - v(x,t)$. 
As a matter of fact, nonlinear effects, i.e. saturation of the perturbation 
growth, are present only in the central part of the disturbance.

We consider now the spatial shape of the leading edge of the 
front and its temporal evolution in the limit $t \to \infty$
for the ideal case of an infinite lattice ($L =\infty$). As a
starting point, we make the {\it ansatz} that for $|x|\gg 1$
\be
\delta u(x,t) =  {\phi(x,t)\over |x|^{\beta}} \quad ,
\label{pow_dis}
\ee
with $\phi(x,t)$ weakly dependent on $x$. Inserting Eq.~(\ref{pow_dis}) in 
Eq.~(\ref{linmap}), we obtain
\be
\frac{\phi(x,t+1)}{|x|^\beta} = m(x,t)
\left[  (1-\varepsilon) \frac{\phi(x,t)}{|x|^\beta}+\varepsilon
\sum_{y\ne x} w(x-y) \frac{\phi(y,t)}{|y|^\beta} 
\right]\quad ,
\label{lm2}
\ee
Due to the long range coupling we can assume that, as a first approximation,
a mean field description holds. This amounts to neglect the spatial 
fluctuations of $\phi(x,t)$ and to replace $m(x,t)$ with its average 
$e^\lambda$, where $\lambda$ is the (maximal) Lyapunov exponent. 
Eq.~(\ref{lm2}) is then splitted in two equations, one for the time 
evolution
\be
\phi(t+1) = e^\lambda \phi(t)
\label{timeg}
\ee 
and one for the spatial profile
\be
\frac{1}{|x|^\beta} = 
\sum_{y\ne x} \frac{w(x-y)}{|y|^\beta} \quad .
\label{spat_profile}
\ee
Moreover, at least for $x\gg 1$, the sum appearing in 
Eq.~(\ref{spat_profile}) can be approximated by neglecting 
the small terms coming from decaying tails, i.e. by extending the
sum only between 1 and $x-1$ (for simmetry reasons, we can also consider
$x>0$). The spatial shape of the front is thus conserved in time if
the fixed-point condition 
\be
\frac{1}{x^\beta} \approx \sum_{0<y<x} \frac{c(\alpha)}{y^\beta (x-y)^\alpha}
\quad ,
\label{fixpoint}
\ee
is satisfied for $x$ large enough. Since the leading contributions 
to the sum came from the extrema and can be estimated to be 
of order $|x|^{-\beta}$ and $|x|^{-\alpha}$, Eq.~(\ref{fixpoint}) is 
fulfilled for $\beta\le\alpha$.
By combining this result with that of Eq.~(\ref{timeg}), we get the 
expression for the asymptotic behaviour of the front leading edge 
\be
\delta u(x,t) \sim { \exp(\lambda t) \over |x|^{\beta} }\quad .
\label{profile}
\ee

Let us define the front position $r(t)$ as the maximal distance 
from $x=0$ where $|\delta u(x,t)|~\ge~\theta$, with $\theta>0$ being a 
preassigned threshold. According to Eq.~(\ref{profile}), $r(t)$ grows 
exponentially as
\be
r(t) \sim \exp{\left(\frac{\lambda}{\beta} t \right)} \equiv \exp
\left(S(\beta) t\right)
\label{exp_law}
\ee
for $t\to \infty$, where the value of the $\beta$ parameter depends on the
initial shape of the disturbance. From the above reported arguments
it is clear that, if the perturbation decays initially as 
$u_0(x) \sim |x|^{-\beta}$, with $\beta \le \alpha$, than its profile
will be preserved during the time evolution. For more general 
initial conditions the selected $\beta$ can be determined by means
of the following argument. Let us consider $u_0(x)$ to be a superposition
of several profiles, each one decaying as $|x|^{-\beta}$, but with 
different $\beta<\alpha$. In the linear approximation, each one of these 
components will propagate independently with a different rate $S(\beta)$.
On the basis of general arguments \cite{vansaar}, we expect that 
(for $u_0(x)$ sufficiently localized) the profile with slowest 
growth rate $S(\alpha)=\lambda/\alpha$ will be selected.

These results are analogous to those found for CML with nearest-neighbours
coupling \cite{disturb} and for the complex Ginzburg-Landau equation 
\cite{cgle}, the main difference being that the tails of the front 
are exponential for the latter case.

We numerically tested the above predictions for a lattice of coupled tent maps
($f(z)=1-2|z|$), with several values
of the coupling constant and chain lengths ranging from $L=1001$ to $L=20001$. 
In particular, we have considered a single-site perturbation $u_0(x)=\delta_{x,0}$, 
where $\delta_{x,y}$ is the usual Kroneker delta.
We computed the time evolution of $\delta u(x,t)$ according to 
Eq.~(\ref{linmap}) and averaged it over different reference trajectories. 
As shown in Fig.~1, the perturbation profile decays on average
as $|x|^{-\alpha}$ while growing exponentially in time. We have also 
verified that an initial disturbance with a decaying profile $|x|^{-\beta}$ is 
conserved only for $\beta \le \alpha$, otherwise a power law decay 
exponent $\alpha$ is always found. Finally, we checked the validity of 
Eq.~(\ref{timeg}) by verifying that, for increasing $t$, the local growth 
rate on the tails of $\delta u$ approaches the Lyapunov exponent. 

A direct numerical test of Eq.~(\ref{exp_law}) is complicated by the presence
of finite-size effects (see however Ref.~\cite{zakopane} for results on
a closely related CML model). An indirect check is accomplished by considering 
a suitably defined comoving Lyapunov exponent $\Lambda$ \cite{deissler}. 
For CML with local coupling, $\Lambda(v)$ is defined
as the asymptotic growth rate of a disturbance in a reference 
frame moving along the ``world line'' $x_v(t)=vt$, where $v$
is the frame velocity ($|v|\le 1$). This amounts to assume 
$\delta u(x_v(t),t) \sim \exp\left (\Lambda(v) t\right) $. Within this scheme, 
the condition $\Lambda(v)=0$ defines the propagation speed of an initially 
localized disturbance \cite{deissler,disturb}.

In the present case, since the long range coupling leads to exponentially 
fast propagation (see Eq.~(\ref{exp_law})), we rather define
comoving Lyapunov exponent $\Lambda(R)$ in a reference frame moving 
along the ``world line" $x_R(t) = [\exp(Rt)-1]$. Therefore, on the leading edge
and for sufficiently long time $t$, the relation
\be
\delta u(x,t) \sim \exp\left(\Lambda(R) t\right) = \exp\left(\lambda t -
\alpha \log x \right)
\label{comovi2}
\ee
should hold. From this equation it is readily seen that the comoving exponent 
must be a linear function of the rate $R$, namely
\be
\Lambda(R) = \lambda - \alpha R \quad .
\label{lineare}
\ee
Notice that, in analogy with system with local coupling, the condition 
$\Lambda(R)=0$ gives exactly the growth rate $S(\alpha)= \lambda/\alpha$,  
in agreement with the above prediction. 

As before, the finite size of the system prevents the numerical computation
of $\Lambda$ for asymptotically large times as requested by its very 
definition. Therefore, we computed its finite-time value 
\be
\Lambda(R,t) = {1\over t}\left\langle \log\left|
{\delta u(x_R(t),t) \over \delta u(0,0)} \right|\right\rangle
\label{stima}
\ee
where $R = \log|x+1|/t$ and $ \langle \cdot \rangle $ is the average
over different reference trajectories. As can be easily realized, 
the maximal accessible rate will decrease as $\log L/t$. Therefore if the 
iteration time is doubled, the chain length should be increased by a factor 
$L$ in order to achieve the same maximal $R$. Due to the large of amount of 
CPU time required by the iteration, it is thus not feasible to consider system 
sizes larger than $10^4$, and the accessible ranges of $R$ and $t$ values are 
limited by this constraint. Nevertheless, the results reported in Fig.~2 
confirm that a linear behaviour $\Lambda(R,t)=\lambda^*(t) - \alpha R$ is 
observed at each time and for $R$ not too small, but with an intercept
$\lambda^*\ne\lambda$ due to the finite time of observation. Empirically, we 
found that this intercept converges according to the rule
$\lambda^*(2 n t) -\lambda^*(t) \simeq {\rm constant}/q^{n-1}$, for some 
$q > 1$, so that we can extrapolate 
its asymptotic value $\lambda^*(\infty)$ on the basis of the available data. 
Indeed, for the situation reported in Fig.~2 ($\alpha=3/2$ and 
$\varepsilon = 1/3$) we have estimated a value $\lambda^*(\infty)=0.335$, 
in excellent agreement with the corresponding Lyapunov exponent 
$\lambda = 0.338$.

The deviation from Eq.~(\ref{lineare}) at small values of $R$ is due
to transient effects.  We have numerically observed that, 
for increasing $t$, the interval of $R$ values where deviations are 
observed reduces.

In conclusion, we have fully identified the mechanism that rules the 
disturbance propagation for systems with power law long range couplings.
This gives at any position in 
space an exponential increase in time, and a power law fall-off with
$x$. The power with which the perturbation decays, is, for generic
initial conditions equal to the power describing the interaction fall-off.
Moreover, the time needed for the disturbance to propagate with finite
amplitude at a given distance $l$ is inversely proportional to the 
Lyapunov exponents and increases logarithmically with $l$.

\acknowledgements{
We are indebdted to P. Grassberger for effective suggestions
and a careful reading of this letter. We thank R. Livi, A. Pikovsky,
A. Politi and O. Rudzick for useful discussions as well as the ISI Foundation 
(Torino) and the EU HC\&M Network ERBCHRX-CT940546 for partial support. 
One of us (A.T.) gratefully acknowledges the European Community 
for the research fellowship No ERBCHBICT941569 and M. Frese, who 
encouraged him to carry on this research.
}

\begin{figure}
\caption{Plot of the logarithm of the average disturbance amplitude 
$\langle|\delta u(x,t)|\rangle$ versus $\ln x$ at three different times 
for coupled tent maps with $\alpha =2$, $L=5001$, $\varepsilon=1/3$. 
From bottom to top the three solid curves correspond to $t=20,40$ and 
60, respectively. The dashed line has a slope 2.}
\end{figure}

\smallskip

\begin{figure}
\caption{The finite-time comoving Lyapunov exponents $\Lambda(R,t)$ as a
function of rate $R$ at time $t = 20,40 $ and 80 (from bottom to top), 
for coupled tent maps with $\alpha =3/2$, $L=5001$, $\varepsilon=1/3$. 
The dashed line represents the asymptotic expression 
$\Lambda(R)=\lambda- 3/2 \cdot R$.}
\end{figure}

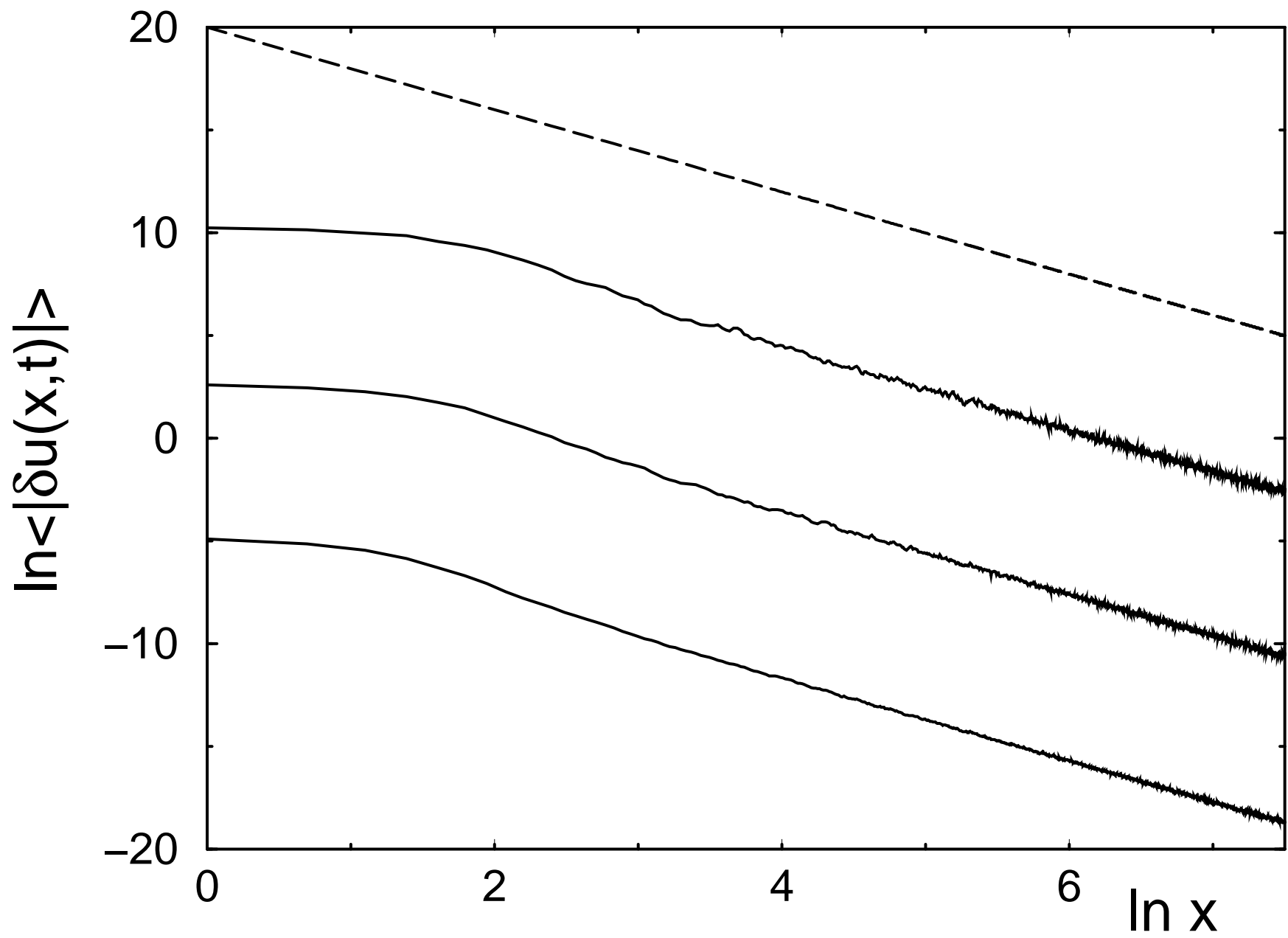
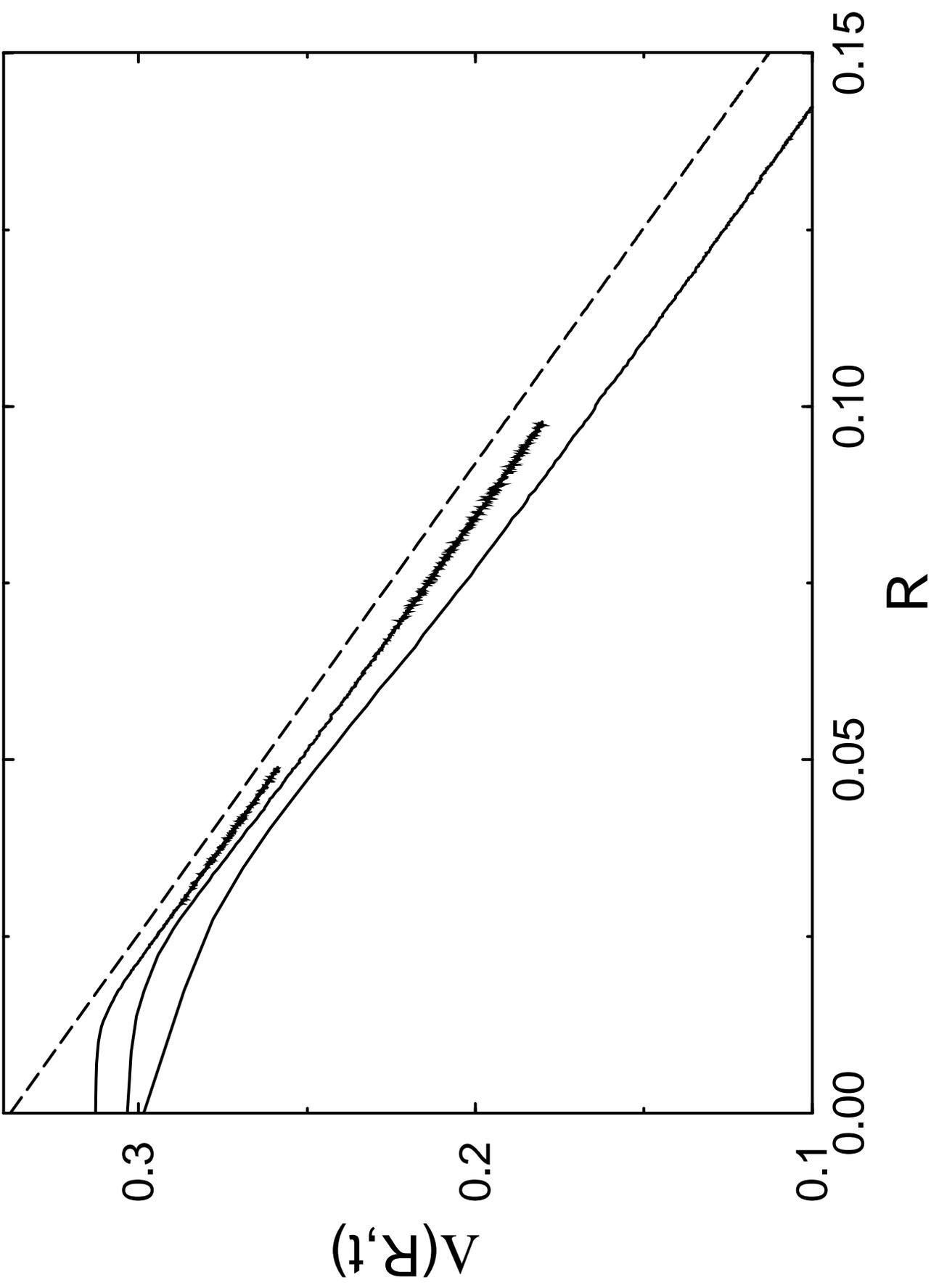
\end{document}